# A Cosmic Ray Resolution to the Superbubble Energy-Crisis

Yousaf M. Butt[1], Andrei M. Bykov[2]

[1]Harvard-Smithsonian Center for Astrophysics, Cambridge, Massachusetts 02138, USA. ybutt@cfa.harvard.edu

[2]A.F. Ioffe Institute of Physics and Technology, St. Petersburg, Russia, 194021
byk@astro.ioffe.ru

## ABSTRACT

Superbubbles (SBs) are amongst the greatest injectors of energy into the Galaxy, and have been proposed to be the acceleration site of Galactic cosmic rays. They are thought to be powered by the fast stellar winds and powerful supernova explosions of massive stars in dense stellar clusters and associations. Observations of the SB "DEM L192" in the neighboring Large Magellenic Cloud (LMC) galaxy show that it contains only about one-third the energy injected by its constituent stars via fast stellar winds and supernovae. It is not yet understood where the excess energy is going – thus, the so-called "energy-crisis".  We show here that it is very likely that a significant fraction of the unaccounted for energy is being taken up in accelerating cosmic rays, thus bolstering the argument for the SB origin of cosmic rays.

# INTRODUCTION

Superbubbles (SBs) are amongst the greatest injectors of energy into the Galaxy (eg. Oey 2004), and have been proposed to be the acceleration site of Galactic cosmic rays (eg. Higdon and Ligenfelter 2005; Parizot 2004; Bykov 2001; Bykov and Fleishman 1992). They are thought to be powered by the fast stellar winds and powerful supernova explosions of massive stars in dense clusters called OB associations – such stellar clusters can contain many hundreds, or even thousands, of massive stars. Observations of the SB "DEM L192" in the neighboring Large Magellenic Cloud (LMC) galaxy show that it contains only about one-third the energy injected by its constituent stars via fast stellar winds and supernovae (Cooper et al., 2004). The excess energy must be going somewhere, but it is not yet understood where – thus, the so-called "energy-crisis". We show here that it is very likely that a significant fraction of the unaccounted for energy is being taken up in accelerating cosmic rays, thus bolstering the argument for the SB origin of cosmic rays (e.g. Binns et al. 2007).

## The DEM L192 Superbubble in the LMC

DEM L192 (also known as N51D) is a ~135 pc by ~120 pc SB in the LMC hosting two OB associations: LH 51 and LH 54 both about ~3Myrs old (Oey & Smedley 1998). In a detailed multiwavelength study, Cooper et al (2004) found that the sum of the thermal and kinetic energy of the SB is $(6 \pm 2) \times 10^{51}$ ergs. In comparison, the total stellar wind energy injected over the lifetime of the host OB associations is $(5 \pm 1) \times 10^{51}$ ergs; in addition, 13±4 supernovae (SN), with a mechanical energy of ~$10^{51}$ ergs each, have exploded in the region yielding another $(13 \pm 4) \times 10^{51}$ ergs of energy. Thus the energy injected into the SB exceeds the observed SB energy by a factor of ~3. DEM L192 is hardly unique in this respect. Most other SBs where such measurements are possible show similar discrepancies: eg. Cyg OB1 (Saken et al., 1992), Orion OB1 (Brown, Hartmann & Burton, 1995), N11B, N108B (Nazé et al. 2001) all seem to be significantly less energetic than their host stellar populations would indicate [see also, eg. Oey (1996)].

What is the unaccounted for energy sink? Various ideas have been put forth: for instance, a superbubble "blowout" – in which the hot gas from the SB interior can spew out – could directly reduce the thermal energy content, but no such blowout is evident in DEM L192 (Cooper et al., 2004). The ambient density or pressure may be underestimated slowing the SB expansion (eg. Garcia-Segura & Oey 2004), or the hot SB interior may be losing more energy than suspected via radiative cooling. [However, X-ray observations do not appear to support the latter scenario (eg. Chu et al., 1995)]. Lastly, the input energy from stellar winds and SNs may be overestimated. While several of these effects may be simultaneously contributing to the discrepancy in SB dynamics we focus here on a different possibility: that significant SB energy may be being transferred to accelerating cosmic rays to high energies. We argue that this energy sink, and the related energy stored in CRs and Magneto Hydrodynamic (MHD) turbulence, explains the bulk of the SB energy budget discrepancy.

Powerful stellar winds and SN produce shocks of different strengths that pass through most of the SB heating the gas, producing turbulence and accelerating CRs.

# The Superbubble CR efficiency Calculation for DEM L192

We have calculated how efficiently the DEM L192 SB may be accelerating CRs based on a model that includes a mixture of turbulent (Fermi-II) and diffusive shock (Fermi-I) acceleration and is tailored to the dimensions of DEM L192 (about 120 pc). The energy gain of the suprathermal nuclei occurs due to large scale (non-resonant) compressive MHD motions of the highly conductive magnetized plasma. [This model is described in detail in Bykov (2001), and references therein.] The model assumes a Gaussian source of large scale turbulent motions in a scale $k_o$ with $\mathcal{E}_k \sim <u^2> exp[(-k/k_o)^2]$. In the particular simulation for DEM L192, we used sqrt$\{<u^2>\}$= 400 km/s, and $2\pi/k_o$ =10 pc. Nonthermal particle injection into the acceleration process was parameterized by a dimensionless parameter $\zeta_e$ – the fraction of shock ram energy initially injected as monoenergetic protons of kinetic energy equal to that of protons at the shock upstream flow. The model accounts for a backreaction of accelerated particles on the large scale motions through energy conservation equations, and for CR escape from the system.

The CR acceleration efficiency of DEM L192 (i.e. the fraction of energy of large scale motions transferred to CRs) is illustrated in Fig. 1. At early evolution stages (the increasing portion of the curves in Fig. 1), CR acceleration occurs in the test particle regime. After reaching the maximal efficiency it drops down because of lower acceleration rate and CR escape. For widely accepted values of the injection parameter, $\zeta_e \sim 0.001$, maximal efficiency of CR acceleration corresponds to an age of 3 million years. In general, we find that at some intermediate evolution stage (roughly a few million years age), the large scale shocks and MHD turbulence in such a SB is an efficient method of accelerating CRs, taking up to approximately a third of the input stellar and SN mechanical energy.

Furthermore, the unthermalized turbulent plasma motions are themselves a reservoir of energy that is unaccounted for in the SB energy budgets and this further assists reconciling the "energy crisis". Indeed, Cooper et al. (2004) have found that a non-thermal X-ray component is needed to properly model the spectrum of the diffuse X-ray emission of DEM L192. Similarly, Bamba et al. (2004) detect diffuse non-thermal X-rays from the "30 Doradus C" SB, also in the LMC.

## CONCLUSIONS

Thus, it appears that the dominant reason that SBs do not contain as much thermal and kinetic energy as is released by their constituent stars and SNs is because most of this energy is taken up in accelerating CRs, as well as being resident in turbulent plasma motions. As such, CR acceleration, and the associated MHD turbulence provides a natural resolution to the SB "energy crisis".

### Acknowledgments

YMB acknowledges partial support from a NASA Long Term Space Astrophysics Grant. AMB was partly supported by RBRF grant 06-02-16844 and by a program of RAS Presidium.

# DEM L192 Superbubble

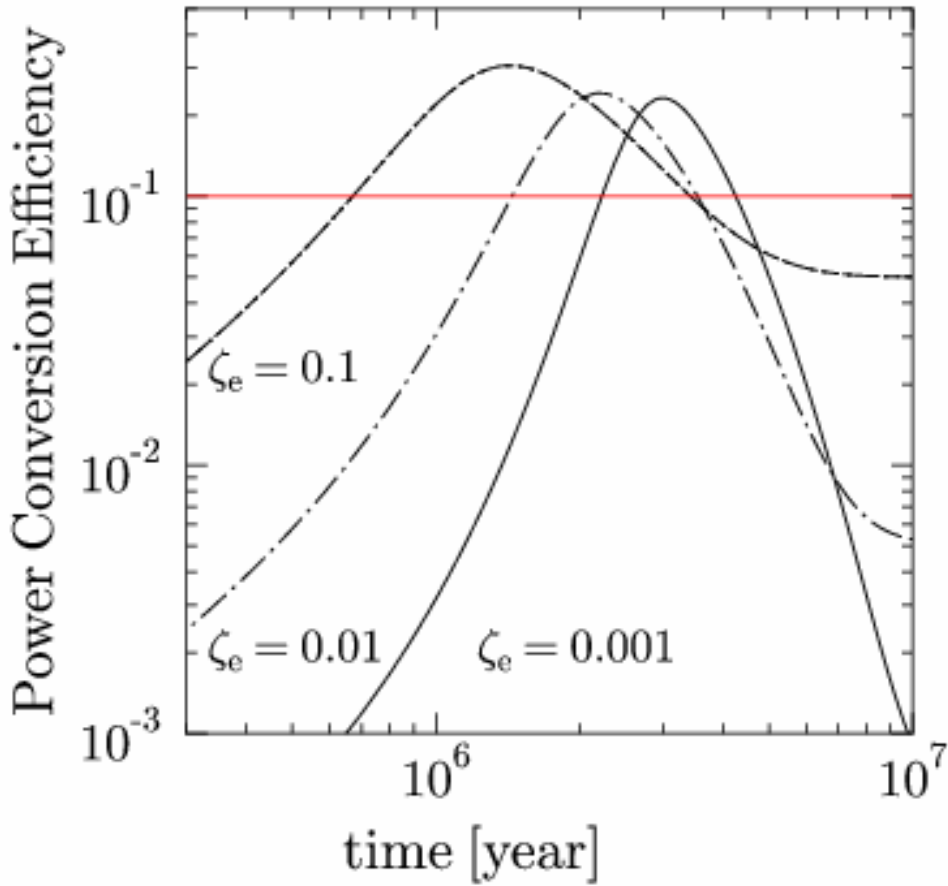

Fig 1. Plotted is the large scale MHD turbulence power conversion efficiency to cosmic ray (CR) acceleration as a function of time, for different CR injection parameters, $\zeta_e$, as calculated for the DEM L192 superbubble (SB). The parameter $\zeta_e$ is the fraction of shock ram pressure initially injected to Fermi acceleration by shocks and large scale motions in DEM L192 (of scale size 120 pc) considered here. As can be seen in the plot, up to a third of the mechanical power of the SB may be taken up by CR acceleration.